\documentclass[conference,letterpaper]{IEEEtran}

\IEEEoverridecommandlockouts

\usepackage[utf8]{inputenc} 
\usepackage[T1]{fontenc}
\usepackage{url}
\usepackage{ifthen}
\usepackage{aliascnt}
\usepackage{hyperref}
\usepackage[cmex10]{amsmath} % Use the [cmex10] option to ensure complicance
\usepackage{amssymb}
\usepackage{amsthm}
\usepackage{graphicx}
\usepackage[style=ieee,isbn=false,url=false,eprint=false]{biblatex}

\usepackage{verbatim}
\usepackage{xcolor}
\usepackage{mathrsfs}
\usepackage[normalem]{ulem}
\usepackage{colortbl}

\newtheorem{theorem}{Theorem}

\newaliascnt{lemma}{theorem}

\aliascntresetthe{lemma}

\newaliascnt{proposition}{theorem}

\aliascntresetthe{proposition}

\newaliascnt{corollary}{theorem}
\newtheorem{corollary}[corollary]{Corollary}
\aliascntresetthe{corollary}

\newaliascnt{example}{theorem}
\newtheorem{example}[example]{Example}
\aliascntresetthe{example}

\newaliascnt{definition}{theorem}
\newtheorem{definition}[definition]{Definition}
\aliascntresetthe{definition}

\newaliascnt{claim}{theorem}

\aliascntresetthe{claim}

\newtheorem{problem}{Problem}

\newtheorem{construction}{Construction}

\renewcommand{\vec}[1]{\ensuremath{\mathbf{#1}}}   % vectors
\newcommand{\cmat}[1]{\ensuremath{#1^{(c)}}}       % Composite matrices
\newcommand{\rll}[2]{\ensuremath{\mathrm{RLL}_{#1}\left({#2} \right) }} % RLL codes
\newcommand{\red}[1]{\ensuremath{\mathrm{red}\left( #1 \right)}} % Redundancy
\newcommand{\prob}[1]{\ensuremath{\mathrm{Pr}\left( #1 \right)}} % Probability

\usepackage{fancyhdr}

\begin{document}
\title{Coding for Strand Breaks in Composite DNA}

%% Many authors with many affiliations:
 \author{%
  \IEEEauthorblockN{\textbf{Frederik~Walter}\IEEEauthorrefmark{1}, 
                    and \textbf{Yonatan~Yehezkeally}\IEEEauthorrefmark{2}}
  \IEEEauthorblockA{\IEEEauthorrefmark{1}%
                    Institute for Communications Engineering, %\\
                    Technical University of Munich, 
                    80333 Munich, Germany}
  \IEEEauthorblockA{\IEEEauthorrefmark{2}%
                    School of Computing, %\\
                    Newcastle University, 
                    Newcastle upon Tyne NE4 5TG, United Kingdom}
  \IEEEauthorblockA{Email: 
                    frederik.walter@tum.de, 
                    yonatan.yehezkeally@ncl.ac.uk}
                    \vspace{-1.5\baselineskip}
   \thanks{Funded by the European Union (DiDAX, 101115134). Views and opinions expressed are however those of the author(s) only and do not necessarily reflect those of the European Union or the European Research Council Executive Agency. Neither the European Union nor the granting authority can be held responsible for them.
   It was also partially funded by UKRI BBSRC grant BB/Y007638/1, the Department for Science, Innovation and Technology (DSIT) and the Royal Academy of Engineering for a Chair in Emerging Technologies award.}%
 }

\maketitle

\pagestyle{empty}
\thispagestyle{fancy}

\begin{abstract}
  Due to their sequential nature, traditional DNA synthesis methods are expensive in terms of time and resources. They also fabricate multiple copies of the same strand, introducing redundancy. This redundancy can be leveraged to enhance the information capacity of each synthesis cycle and DNA storage systems in general by employing composite DNA symbols. Unlike conventional DNA storage, composite DNA encodes information in the distribution of bases across a pool of strands rather than in the individual strands themselves. Consequently, error models for DNA storage must be adapted to account for this unique characteristic. 
  One significant error model for long-term DNA storage is strand breaks, often caused by the decay of individual bases. This work extends the strand-break channel model to the composite DNA setting. To address this challenge, we propose a coding scheme that uses marker codes to correct single strand breaks. As part of this approach, we generalise run-length-limited (RLL) codes for the composite setting and derive bounds on their redundancy.
\end{abstract}

\section{Introduction}

Storing data on DNA molecules presents a promising solution for archiving vast amounts of information due to its high density and long-term stability. Traditional digital storage media, such as hard drives and magnetic tapes, are constrained by physical size and degradation over time. In contrast, DNA, the molecule that carries genetic information in living organisms, provides a compact and durable medium for data storage. This potential has been demonstrated in several pioneering studies \autocite{church_2012_NextGenerationDigital, goldman_2013_PracticalHighcapacity, erlich_2017_DNAFountain, organick_2018_RandomAccess}.

The most used DNA synthesis techniques create thousands to millions of copies of the same strand, which introduces redundancy \cite{heller_2002_DNAMicroarray,bumgarner_2013_OverviewDNA,lietard_2021_ChemicalPhotochemical}.
Recent advancements have proposed the composite DNA synthesis method to leverage this redundancy and increase the information capacity of DNA data storage systems \autocite{anavy_2019_DataStorage}. 
Instead of storing information in individual DNA strands, composite DNA encodes information in the distribution of bases across a pool of strands. This approach allows for encoding more information in a single synthesis process, thereby alleviating a major driver of cost and latency in concurrent synthesis techniques.

The composite DNA symbols are defined by mixtures of DNA bases and their ratios at specific positions in the strands. This effectively extends the alphabet beyond the four standard bases. A composite DNA symbol is represented as a quartet of probabilities $\{p_A, p_C, p_G, p_T\}$, where each $p_X$ denotes the fraction of base $X \in \{A, C, G, T\}$ in the mixture, and $p_A + p_C + p_G + p_T = 1$. Identifying composite symbols requires sequencing multiple reads, synchronising them, and estimating the base fractions at each position. Coding for various models of the composite channel has been proposed in \autocite{zhang_2022_LimitedMagnitudeError,walter_2024_CodingComposite,sabary_2024_ErrorCorrectingCodesa}.

However, storing and retrieving data using DNA is not without challenges. Many of them have been addressed \autocite{heckel_2017_FundamentalLimits,shomorony_2021_DNABasedStorage,lenz_2020_AchievingCapacitya}. 
One relevant error due to long-term storage is strand breaks that occur when the DNA molecule is cleaved into two or more fragments, which can complicate reconstructing the original data \autocite{meiser_2022_InformationDecay,lindahl_1993_InstabilityDecay,allentoft_2012_HalflifeDNA}. The problem of correcting breaks in traditional DNA data storage channels has been addressed in several studies \autocite{shomorony_2021_TornPaperCoding,ravi_2021_CapacityTorn,bar-lev_2023_AdversarialTornpaper}, which propose various coding schemes to mitigate the impact of such errors.
The strand break can be modelled as a random process with an exponential decay rate for each bond in the DNA backbone. The decay rate is mainly influenced by the encapsulation method, water content, temperature, and PH level, leading to a half-life in the range of 30 years up to 158,000 years. \autocite{allentoft_2012_HalflifeDNA,grass_2015_RobustChemical}. Therefore, we can assume a probability for each bond to break \cite{gimpel_2024_ChallengesErrorcorrection} and model the number of breaks as a Poisson process. 

Since the concept of composite DNA uses the redundancy of the synthesis process to increase the information density, we do not have multiple copies of the same strand. Therefore, alignment methods or codes for shotgun sequencing (see, e.g., \autocite{motahari_2013_InformationTheory,ravi_2022_CodedShotgun}) are not applicable. 
This paper addresses the problem of correcting breaks in DNA sequences originating from composite DNA synthesis. 
As a first step in that direction, we propose a coding scheme employing a marker to correct single breaks and introduce a generalisation for run-length-limited codes applicable to the composite setting. Our approach aims to enhance the reliability and efficiency of data retrieval in composite DNA storage systems, ensuring accurate reconstruction of stored information despite strand breaks.

The remainder of this paper is organised as follows. In Section II, we define the problem and state our objectives. In Section III, we introduce and analyse composite run-length-limited (RLL) codes, including bounds on their redundancy. Section IV presents our proposed marker-based coding scheme for correcting single breaks in DNA sequences. Finally, Section V concludes the paper and discusses potential directions for future research.

\section{Definitions and Problem Statement}

For positive integers $k,n \in \mathbb{N}$, let $[k,n] = \{ k, k+1, \dots, n \}$. The binomial coefficient $n$ choose $k$ is denoted by $\binom{n}{k}$. 
Our approach to modelling composite DNA is described as follows. For ease of notation, we normalise the composite letter probabilities to an integer $M \in \mathbb{N}$, which we call the \emph{resolution parameter}, as in \autocite{anavy_2019_DataStorage}. We then define the composite symbol and matrix as follows.

\begin{definition}[Composite Symbol and Matrix]
  \label{def:composite_matrix}
  Let $q>1$ be the size of the base alphabet (in the case of DNA, it is 4), and let the integer $M>0$ be the \emph{resolution parameter}. A \emph{composite symbol} is given by the q-tuple $\mathbf{x} = (x_1, x_2, \ldots, x_q) \in [0,M]^q$ such that $\sum_{i=1}^q x_i = M$. For an integer $n>0$, a matrix $\cmat{X} \in [0,M]^{q \times n}$ is called a \emph{composite matrix} if each column represents a composite symbol. It is therefore the concatenation of $n$ composite symbols. We denote the $j$-th column of $\cmat{X}$ by $\cmat{X}_j$, and the $i$-th entry of the $j$-th column by $\cmat{X}_{i,j}$.
\end{definition}

We use lowercase letters $q$ to represent the base alphabet and uppercase letters $Q$ to represent the size of the composite alphabet, i.e., the number of possible composite symbols. Given a resolution parameter $M$, we have $Q=\binom{M+q-1}{q-1}$.

\begin{figure*}[t]
  \centering
  \includegraphics[width=0.8\textwidth]{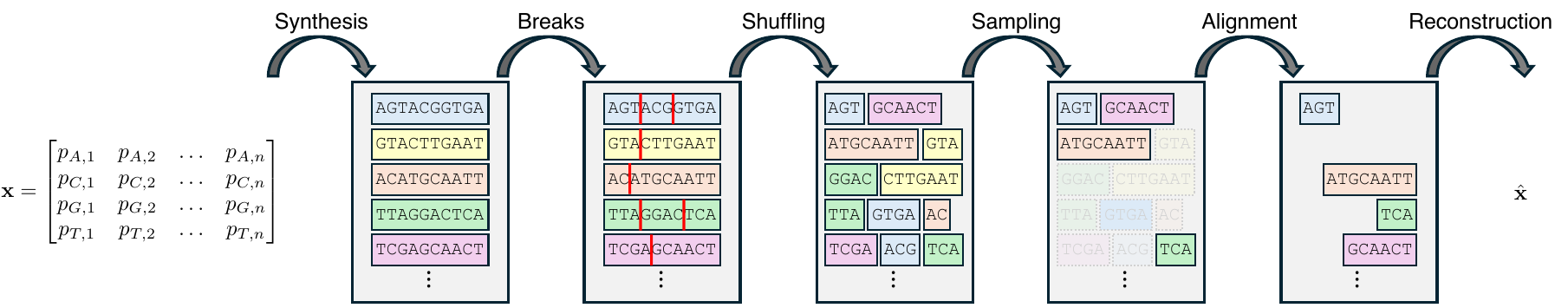}
  \caption{Model for the composite strand break channel.}
  \label{fig:channel_model}
\end{figure*}

The composite matrix represents the probability distribution of the composite DNA pool. As shown in \autoref{fig:channel_model}, the DNA is synthesised into strands with the probability distribution for each position as given in the composite matrix. Due to the large number of strands produced with standard DNA synthesis methods, we can assume that the strands are independent and identically distributed.

Each strand is subject to at most $t$ breaks, resulting in at most $t+1$ fragments per strand, which are unordered. From this pool of DNA, we sample $K$ fragments. The fragments are then sequenced and require alignment. Since the strands originate from composite DNA, we cannot rely on overlapping subsequences, as in standard alignment methods, to determine the correct position of the fragments.

\begin{problem}
  Create a code such that any fragment of a strand resulting from multiple breaks in the DNA strand can be correctly positioned.
\end{problem}

This paper does not consider probabilistic errors originating from the sampling process with repetition. A method to handle this, which is applicable here, can be found in \autocite{zhang_2022_LimitedMagnitudeError}. 
We suggest a marker-based approach for coding for this channel. As this work does not yet present an analysis of the capacity of the composite strand break channel, we cannot study the redundancy of our chosen coding approach in the context of the channel's capacity. Nevertheless, a requisite component in our strategy is ensuring that markers do not appear in the data itself; for this purpose, we study the following problem.

\begin{problem}
  Generalise the concept of Run-Length-Limited (RLL) codes to the composite setting, determine bounds for the code size, and propose a construction.
\end{problem} 

Furthermore, given a composite run-length-limited code, we want to determine the ideal length of the marker sequence in terms of redundancy.

\section{Composite Run-Length-Limited Codes}

As we want to use marker sequences to identify the position of fragments, we must ensure that the marker sequence is not part of the data. To achieve this, we will use run-length-limited (RLL) codes. These codes date back to magnetic media, where they were used to avoid long runs of the same symbol 
\autocite[Sections 1.1, 1.2]{marcus_2001_IntroductionCoding}.

The approach of \autocite{levy_2019_MutuallyUncorrelated} for encoding such sequences, which achieves only one bit of redundancy for avoiding run-lengths of $\log(n)$, is not applicable in the composite setting, as spontaneous combinations of composite symbols can result in the marker sequence in the synthesised strand. 
Therefore, we require a generalisation of run-length-limited codes that applies to the composite setting. We define these codes as follows:

\begin{definition}[Composite RLL Code]
  \label{def:composite_rll_code}
  Given a composite matrix $\cmat{X}$ as in \autoref{def:composite_matrix} of length $n$ with resolution parameter $M$ over a base alphabet of size $q$, we obtain an alphabet $\Sigma$ of size $Q = \binom{M+q-1}{q-1}$ for the composite symbols. Of these composite symbols, let $R$ symbols belong to an alphabet $\Sigma'\subseteq \Sigma$. We say that a composite matrix is \emph{$\ell$-run-length-limited} if every window of $\ell$ consecutive symbols contains a symbol of $\Sigma\setminus\Sigma'$
  , and denote the set of all such matrices by $\rll{Q,R}{\ell, n}$.
\end{definition}

In the following, we will derive bounds on the redundancy of these composite RLL codes. To do so, we will employ counting arguments to estimate the number of undesired sequences.

\begin{theorem}
  \label{the:red_rll_low}
  A lower bound on the redundancy of $\rll{Q,R}{\ell, n}$ is given by
  \begin{align*}
    \red{\rll{Q,R}{\ell, n}} 
    \geq 
    \log_Q(e) \big( \tfrac{R}{Q} \big)^\ell \big( 1-\tfrac{R}{Q} \big) \cdot 
    \tfrac{n - 2\ell}{2}. 
  \end{align*}
\end{theorem}
\begin{IEEEproof}
  To generalise the analysis of \autocite[Lem.~3]{levy_2019_MutuallyUncorrelated}, we partition the sequence into segments of length $\frac{n}{2 \ell}$ and notice that if the whole sequence is $\ell$-run-length-limited, then so is each and every segment.
  Formally: 
  \begin{align}\label{eq:rll-2ell}
    \rll{Q,R}{\ell,n} \subseteq 
    \left(\rll{Q,R}{\ell,2\ell} \right)^{\lfloor \frac{n}{2 \ell}\rfloor} \times 
    \Sigma^{n \bmod 2\ell}
  \end{align}

  We aim to find the size of $\rll{Q,R}{\ell,2\ell}$, or equivalently, the size of $\Sigma^{2\ell} \setminus \rll{Q,R}{\ell,2\ell}$. Note that a sequence of length $2\ell$ cannot contain two \emph{disjoint} runs of length larger than $\ell$ over $\Sigma'$; therefore, we can count the number of undesired sequences by the index $j \in [1, \ell+1]$ where such a run begins, and the length of the run $k \in [\ell,2\ell-j]$. For each corresponding choice of $j,k$, the number $N_{j,k}$ of such sequences is given by 
\begin{align*}
  N_{j,k} = 
  \begin{cases}
    R^{2\ell} & \text{ if } j=1, k=2\ell \\
    R^k Q^{2\ell-k-1} \left( Q-R \right) & \text{ if } j=1, k \leq 2\ell \\
    R^k Q^{2\ell-k-1} \left( Q-R \right) & \text{ if } j>1, k = 2\ell - j +1 \\
    R^k Q^{2\ell-k-2} \left( Q-R \right)^2 & \text{ if } j > 1, k<2\ell-j+1.
  \end{cases}
\end{align*}
 The reasoning for each line is as follows: 1. $2\ell$ symbols from $\Sigma'$ need to be chosen; 2. One chooses $k$ symbols over $\Sigma'$, and a single symbol over $\Sigma\setminus\Sigma'$ for the index $k$, where the remaining $2\ell-k-1$ symbols are chosen freely over $\Sigma$; 3. One chooses $k$ symbols over $\Sigma'$, and a single symbol over $\Sigma\setminus\Sigma'$ for the index $j-1$, where the remaining $2\ell-k-1$ symbols are chosen freely over $\Sigma$; 4. One chooses $k$ symbols over $\Sigma'$, and two symbols over $\Sigma\setminus\Sigma'$ for the indices $j-1$ and $j + k$, where the remaining $2\ell-k-2$ symbols are chosen freely over $\Sigma$. 
Then we get 
\begin{IEEEeqnarray}{+rCl+x*}
  \left\vert \rll{Q,R}{\ell,2\ell} \right\vert &=& Q^{2\ell} - R^{2\ell} - 
  \sum_{j=1}^\ell R^{2\ell-j} Q^j \big(1-\tfrac{R}{Q}\big) \>-
  \nonumber\\ 
  && \sum_{k=\ell}^{2\ell-1} R^k Q^{2\ell-k} \big(1-\tfrac{R}{Q}\big) \>- 
  \nonumber \\ 
  && \sum_{j=1}^{\ell-1} \sum_{k=\ell}^{2\ell-j-1 } R^k Q^{2\ell-k} \big(1-\tfrac{R}{Q}\big)^2. \label{eq:rll-2ell_size}
\end{IEEEeqnarray}
We first note that for the first two summations, we have
\begin{align}
  \sum_{j=1}^\ell & R^{2\ell-j} Q^j \big(1-\tfrac{R}{Q}\big) + 
  \sum_{k=\ell}^{2\ell-1} R^k Q^{2\ell-k} \big(1-\tfrac{R}{Q}\big) 
  \nonumber\\
  &= R^\ell \big(1-\tfrac{R}{Q}\big) \left( \sum_{j=1}^\ell R^{\ell-j} Q^j + 
  \sum_{k=\ell}^{2\ell-1} R^{k-\ell} Q^{\ell + \ell - k} \right) \nonumber\\
  &= 2 R^\ell \big(1-\tfrac{R}{Q}\big) \sum_{i=0}^{\ell-1} R^i Q^{\ell-i} \nonumber\\
  &= 2 (Q R)^\ell \big(1-\tfrac{R}{Q}\big) \sum_{i=0}^{\ell-1} \big( \tfrac{R}{Q} \big)^i \nonumber\\
  &= 2 (Q R)^\ell \left( 1 - \big(\tfrac{R}{Q} \big)^\ell \right). \label{eq:twosums}
\end{align}
Secondly, for the last double sum we get 
\begin{align}
  \sum_{j=1}^{\ell-1}& \sum_{k=\ell}^{2\ell-j-1} R^k Q^{2\ell-k} \big(1-\tfrac{R}{Q}\big)^2 
  \nonumber\\
  &= (RQ)^\ell \big( 1-\tfrac{R}{Q} \big)^2 
  \sum_{j=1}^{\ell-1} \sum_{i=0}^{\ell-j-1} \big( \tfrac{R}{Q} \big)^i \nonumber\\
  &= (RQ)^\ell \big( 1-\tfrac{R}{Q} \big)^2 \sum_{i=0}^{\ell-2} 
  \sum_{j=1}^{\ell-1-i} \big( \tfrac{R}{Q} \big)^i \label{eq:change_summation}\\
  &= (RQ)^\ell \big(1-\tfrac{R}{Q} \big)^2 \sum_{i=0}^{\ell-2} 
  (\ell-1-i) \big( \tfrac{R}{Q} \big)^i \nonumber\\
  &= (RQ)^\ell \bigg[ (\ell-1) \big( 1-\tfrac{R}{Q} \big) \left( 1 - \big(\tfrac{R}{Q}\big)^{\ell-1} \right) \>- \nonumber\\*
  & \hphantom{= (RQ)^\ell \bigg[} \big( 1-\tfrac{R}{Q} \big)^2 \sum_{i=0}^{\ell-2} i \big( \tfrac{R}{Q} \big)^i \bigg] \nonumber\\
  &= (RQ)^\ell \bigg[(\ell-1) \left( 1 - \tfrac{R}{Q} - \big( \tfrac{R}{Q} \big)^{\ell-1} + \big(\tfrac{R}{Q} \big)^\ell \right) \>- \nonumber\\*
  & \hphantom{= (RQ)^\ell \bigg[} \big( 1-\tfrac{R}{Q} \big)^2 \sum_{i=0}^{\ell-2} i \big( \tfrac{R}{Q} \big)^i \bigg], \label{eq:double-sum}
\end{align}
where \eqref{eq:change_summation} is justified by a change in summation order (see Appendix). 
To simplify the last sum, we note 
\begin{align*}
  & \big(1- \tfrac{R}{Q}\big) \sum_{i=0}^{\ell-2} i \big(\tfrac{R}{Q} \big)^i 
  = \sum_{i=0}^{\ell-2} i \big(\tfrac{R}{Q} \big)^i - 
  \sum_{i=0}^{\ell-2} i \big(\tfrac{R}{Q} \big)^{i+1} \\
  &= \sum_{i=0}^{\ell-2} i \big(\tfrac{R}{Q} \big)^i - 
  \sum_{j=1}^{\ell-1} (j-1) \big(\tfrac{R}{Q} \big)^j \\
  &= \sum_{i=1}^{\ell-2} i \big(\tfrac{R}{Q} \big)^i - 
  \sum_{j=1}^{\ell-2} (j-1) \big(\tfrac{R}{Q} \big)^j - 
  (\ell-2) \big(\tfrac{R}{Q} \big)  ^{\ell-1} \\
  &= \sum_{i=1}^{\ell-2} \big(\tfrac{R}{Q} \big)^i - 
  (\ell-2) \big(\tfrac{R}{Q} \big)^{\ell-1}, 
\end{align*}
and therefore 
\begin{align*}
  \big(1-\tfrac{R}{Q} \big)^2 & \sum_{i=0}^{\ell-2} i \big(\tfrac{R}{Q} \big)^i 
  \\
  &= \tfrac{R}{Q} - \big(\tfrac{R}{Q} \big)^{\ell-1} - 
  (\ell-2) \big( 1-\tfrac{R}{Q} \big) \big(\tfrac{R}{Q} \big)^{\ell-1} \\
  &= \tfrac{R}{Q} - (\ell-1) \big(\tfrac{R}{Q} \big)^{\ell-1} + 
  (\ell-2) \big(\tfrac{R}{Q} \big)^\ell.
\end{align*}
It follows by substituting back into \eqref{eq:double-sum} that 
\begin{align}
  \sum_{j=1}^{\ell-1} \sum_{k=\ell}^{2\ell-j-1} & R^k Q^{2\ell-k} \big(1-\tfrac{R}{Q}\big)^2 
  \nonumber\\
  =&\> (Q R)^\ell \bigg[(\ell-1) \left(1 - \tfrac{R}{Q} - \big(\tfrac{R}{Q}\big)^{\ell-1} + \big(\tfrac{R}{Q}\big)^\ell \right) \>- \nonumber\\*
  &\> \hphantom{(Q R)^\ell \bigg[} 
   \tfrac{R}{Q} + (\ell-1) \big(\tfrac{R}{Q}\big)^{\ell-1} - 
  (\ell-2) \big(\tfrac{R}{Q}\big)^\ell \bigg] \nonumber\\
  =&\> (Q R)^\ell \left( (\ell-1) - \ell \big(\tfrac{R}{Q}\big) + \big(\tfrac{R}{Q}\big)^\ell \right). \label{eq:double-sum_final}
\end{align}
Summing up, we have from \eqref{eq:rll-2ell_size}, \eqref{eq:twosums} and \eqref{eq:double-sum_final} that
\begin{IEEEeqnarray}{+rCl+x*}
  \IEEEeqnarraymulticol{3}{l}{%
  \left\vert \rll{Q,R}{\ell,2\ell}\right\vert} 
  \nonumber\\
  &=& Q^{2\ell} - R^{2\ell} - 
  2 (Q R)^\ell \left( 1 - \big(\tfrac{R}{Q}\big)^\ell \right) \>-
  \nonumber\\
  && (Q R)^\ell \left( (\ell-1) - \ell \big(\tfrac{R}{Q}\big) + \big(\tfrac{R}{Q}\big)^\ell \right) 
  \nonumber\\
  &=& Q^{2\ell} \left( 1 - \big(\tfrac{R}{Q}\big)^\ell 
  \big( (\ell+1) - \ell \big(\tfrac{R}{Q}\big) \big) \right).
  \label{eq:rll-2ell_size_final}
\end{IEEEeqnarray}

Finally, from \eqref{eq:rll-2ell} we can now derive 
\begin{align*}
  & \left\vert \rll{Q,R}{\ell,n} \right\vert 
  \\
  &\leq \left[Q^{2\ell} \left(1 - \big(\tfrac{R}{Q}\big)^\ell 
  \big((\ell+1) - \ell \big(\tfrac{R}{Q}\big)\big) \right) \right]^{\left\lfloor n/2\ell \right\rfloor} 
  Q^{n \bmod 2\ell} 
  \\
  &= Q^n \left[1 - \big(\tfrac{R}{Q}\big)^\ell 
  \big(1 + \ell \big(1-\tfrac{R}{Q}\big)\big) \right] ^{\left\lfloor n/2\ell \right\rfloor }. 
\end{align*}
It follows that 
\begin{IEEEeqnarray*}{+rCl+x*}
  && \red{\rll{Q,R}{\ell,n}} 
  \\
  &&\geq n - \log_Q \left[Q^n \left[1 - \big(\tfrac{R}{Q}\big)^\ell 
  \left(1 + \ell \big(1-\tfrac{R}{Q}\big)\right) \right] ^{\left\lfloor n/2\ell \right\rfloor } \right] 
  \\
  &&= - \left\lfloor n/2\ell \right\rfloor \log_Q \left[1 - (\tfrac{R}{Q})^\ell 
  \left(1 + \ell \big(1-\tfrac{R}{Q}\big)\right) \right] \\
  &&\geq \log_Q(e) \big( \tfrac{R}{Q} \big)^\ell \big( 1-\tfrac{R}{Q} \big) \cdot 
  \tfrac{n - 2\ell}{2}, 
\end{IEEEeqnarray*}
where the last inequality follows from $-\ln(1-x)\geq x$.
\end{IEEEproof}

Next, we shall establish an upper bound on the redundancy. 
Before commencing an analysis, we acknowledge the trivial bound 
\begin{align}
  \red{\rll{Q,R}{\ell,n}} 
  &\leq n - \log_Q\left((Q-R)^{\left\lfloor{n/\ell} \right\rfloor } Q^{n - \left\lfloor n/\ell \right\rfloor} \right) 
  \nonumber\\
  &= -\left\lfloor \tfrac{n}{\ell} \right\rfloor \log_Q \big( 1-\tfrac{R}{Q} \big) 
  \nonumber \\ 
  & = \left\lfloor \tfrac{n}{\ell} \right\rfloor \log_Q \big( \tfrac{Q}{Q-R} \big), 
  \label{eq:red-rll_triv}
\end{align}
derived from the size of 
\begin{align*}
\left\{ \vec{x} \in \Sigma^n \mkern-5mu : i \equiv (\ell-1) \mkern-15mu \pmod\ell \Rightarrow x_i\not\in \Sigma' \mkern-5mu \right\} \subseteq \rll{Q,R}{\ell,n}. 
\end{align*}
This bound may prove useful as a baseline in finite-length analysis.

\begin{theorem}
  \label{the:red_rll_up}
  The redundancy of an $\rll{Q,R}{\ell,n}$ code is upper bounded by
  \begin{align*}
    \red{\rll{Q,R}{\ell,n}} 
    \leq 
    e \log_Q(e) \! \big( \tfrac{R}{Q} \big)^\ell \! \left( 1 \! + \! \big( 1 - \tfrac{R}{Q} \big) \! (n-\ell) \right) \! .
  \end{align*}
\end{theorem}
\begin{IEEEproof}
  Denote the probability to draw an $\rll{Q,R}{\ell, n}$ sequence from a uniform distribution to $\Sigma^n$ as $\prob{\rll{Q,R}{\ell, n}}$. Then we have 
  \begin{align*}
    \red{\rll{Q,R}{\ell,n}} = - \log_Q \prob{\rll{Q,R}{\ell,n}}.
  \end{align*}
  Furthermore, we can represent the code as 
  \begin{align*}
    \rll{Q,R}{\ell, n} = \Sigma^n \setminus \bigcup_{i=1}^{ n - \ell +1} \mathcal{A}_i ,
  \end{align*}
  where $\mathcal{A}_i$ is the set of all sequences $ \mathbf{x}$ that contain a substring of length $\ell$ or larger over the alphabet $\Sigma'$ starting at position $i$. That is $\mathbf{x}_{i-1} \notin \Sigma'$ and $\mathbf{x}_{i+k} \in \Sigma'$ for all $ 0 \leq k \leq \ell-1$.
  We define the probability of drawing a sequence from $\mathcal{A}_i$ as $\pi_1 = \big( \frac{R}{Q} \big)^\ell$, and $\pi = \big( \frac{R}{Q} \big)^\ell \big( 1- \frac{R}{Q} \big) $.  
  From the union bound, we get 
  \begin{align*}
    \prob{\rll{Q,R}{\ell,n}} 
    &\geq 1 - \sum_{i=1}^{n-\ell +1} \prob{\mathcal{A}_i} 
    \\
    &= 1 - \pi_1 - (n - \ell) \pi 
    \\
    &= 1 - \big( \tfrac{R}{Q} \big)^\ell \! - (n-\ell) \! \big( 1- \tfrac{R}{Q} \big) \! \big( \tfrac{R}{Q} \big)^\ell 
    \\
    &= 1 - \big( \tfrac{R}{Q} \big)^\ell \left( 1 + \big( 1- \tfrac{R}{Q} \big) (n-\ell) \right).
  \end{align*}
  When this lower bound is positive, it implies that 
  \begin{align}
    &\red{\rll{Q,R}{\ell,n}} 
    \nonumber\\
    &\leq - \log_Q \left( 1 - \big( \tfrac{R}{Q} \big)^\ell \left( 1 + \big( 1- \tfrac{R}{Q} \big) (n-\ell) \right) \right) 
    \nonumber\\
    &\leq \log_Q(e) \frac{ \big( \tfrac{R}{Q} \big)^\ell \left( 1 + \big( 1- \tfrac{R}{Q} \big) (n-\ell) \right)}{1 - \big( \tfrac{R}{Q} \big)^\ell \left( 1 + \big( 1- \tfrac{R}{Q} \big) (n-\ell) \right)} \label{eq:lb_union_bound}.
  \end{align}
  We can replace the union bound with Lovasz's local lemma \autocite[Th. 1.1]{spencer_1977_AsymptoticLower} as in \autocite[Th. 4, Cor. 5]{yehezkeally_2024_CodesNoisy}.  
  Define $\Gamma_i = \left\{ \mathcal{A}_j : \left\vert i-j \right\vert < \ell + 1 \right\}$, and we see that the events $\mathcal{A}_i$ are independent of the events $\left\{ \mathcal{A}_j \notin \Gamma_i \right\}$. In addition, note that there are at most $ 2 \ell + 2$ events in $\Gamma_i$, of which at most one is $\mathcal{A}_1$.
  From \autocite[Cor. 5]{yehezkeally_2024_CodesNoisy} we have that if for all $i$ there exists constants $0< \phi_i<1$ such that
  \vspace{-1em}
  \begin{align*}
    \prob{\mathcal{A}_i} \leq \phi_i \exp \left( - \sum_{j \in \Gamma_i} \phi_j - \phi_i \right)
  \end{align*}
  Then we have 
  \vspace{-0.5em}
  \begin{align}
    \prob{\Sigma^n \setminus \bigcup_{i=1}^{ n - \ell +1} \mathcal{A}_i} \geq \exp \left( - \sum_{i=1}^{n-\ell +1} \phi_i \right). \label{eq:local_lemma}
  \end{align}
  We define $\phi = e \pi $ and $ \phi_1 = e \pi_1 $. If 
  \begin{align*}
    \pi &\leq \phi \exp \left( -(2 \ell +2)\phi - \phi_1 \right) 
    \\
    & = \pi \exp \left( 1 - e ((2 \ell + 2 ) \pi + \pi_1) \right)
  \end{align*}
  and
  \begin{align*}
    \pi_1 &\leq \phi_1 \exp \left( -(\ell -1)\phi - \phi_1 \right) 
    \\
    & = \pi_1 \exp \left( 1 - e ((\ell -1 ) \pi + \pi_1) \right),
  \end{align*}
  which holds for sufficiently large~$\ell$, by \eqref{eq:local_lemma} we get
  \begin{align*}
    \prob{\rll{Q,R}{\ell,n}} 
    &\geq \exp \left( - \phi_1 - (n- \ell) \phi \right) 
    \\
    &= \exp \left( -e (\pi_1 + (n-l) \pi) \right).
  \end{align*}
  Put differently,
  \begin{align*}
    \red{\rll{Q,R}{\ell,n}} 
    &\leq e \log_Q(e) (\pi_1 + (n-\ell) \pi)
    \\
    &= e \log_Q(e) \big( \tfrac{R}{Q} \big)^\ell \left( 1 + \big( 1- \tfrac{R}{Q} \big) (n-\ell) \right).
  \end{align*}
  This holds for all sufficiently large~$n$. However, if $\big( \tfrac{R}{Q} \big)^\ell n $ is small, we get a tighter bound from \eqref{eq:lb_union_bound}.
\end{IEEEproof}

\section{Code Construction}

If we assume that the strands encounter at most one break, the fragments will be either prefixes or suffixes of the original strands. In this case, we can use marker codes to determine the correct position of the fragments. We will add a marker sequence of length $\ell+2$ at the beginning and at the end to identify the position. The marker sequence for a $q$-ary base alphabet will be of the form $ (1, 0, \ldots , 0 , 1) $ with $ \ell $ zeros between the ones. 
Positioning markers at both the beginning and end of the sequences is necessary to avoid confusing a shorter run of zeros in a sampled fragment as a partial occurrence of the marker. For the composite setting, this means that the marker sequence is the same as in the following example. 

\begin{example}
  \label{ex:composite_marker}
  Let $\cmat{X}$ be a composite matrix as in \autoref{def:composite_matrix} over an alphabet of size $q$. Then the composite matrix with the marker sequence is given by: 
  \begingroup
  \setcounter{MaxMatrixCols}{20}
  \begin{align*}
    \footnotesize
    \cmat{X} = 
    \setlength\arraycolsep{2pt}
    \begin{bmatrix}
      0 & M & \cdots & M & 0 & x_{1,\ell +3} & \cdots & x_{1,n - \ell -2 } & 0 & M  & \cdots & M & 0 \\
      M & 0 & \cdots & 0 & M & x_{2,\ell +3} & \cdots & x_{2,n - \ell -2 } & M & 0 & \cdots & 0 & M  \\
      0 & 0 & \cdots & 0 & 0 & x_{3,\ell +3} & \cdots & x_{3,n - \ell -2 } & 0 & 0 & \cdots & 0 & 0  \\
      \vdots & \vdots & \ddots & \vdots & \vdots & \vdots & \ddots & \vdots & \vdots & \vdots & \ddots & \vdots & \vdots \\
      0 & 0 & \cdots & 0 & 0 & x_{q,\ell +3} & \cdots & x_{q,n - \ell -2 } & 0 & 0 & \cdots & 0 & 0 
    \end{bmatrix}
  \end{align*}
  \endgroup 
\end{example}

Observe that the marker sequence results in classical, non-composite symbols in the synthesised strands. This allows for the determination of the position for every segment individually, without relying on the combination of multiple fragments. 
Having introduced composite marker sequences and composite RLL codes, we now present a practical code construction assuming that most strands experience at most one break.

We focus on cases where a strand undergoes approximately one break, resulting in fragments that are either prefixes or suffixes of the original strand. If a fragment is neither a prefix nor a suffix—indicating that the strand has undergone more than one break—it will be discarded.

To identify prefixes and suffixes, we append a marker sequence of length $\ell + 2$ to each strand's beginning and end. The marker sequence is constructed in \autoref{ex:composite_marker}. Since the problem of designing an RLL encoder achieving the redundancy bound in \autoref{the:red_rll_up} remains open, we currently use a practical encoder relying on breaker symbols as in \eqref{eq:red-rll_triv} to ensure that the marker sequence does not appear in the data of the synthesized strands (any encoder developed in the future can be swapped in, if advantageous). 
This leads to the following code construction.

\begin{figure}[t]
  \centering
  \includegraphics[width=0.38\textwidth]{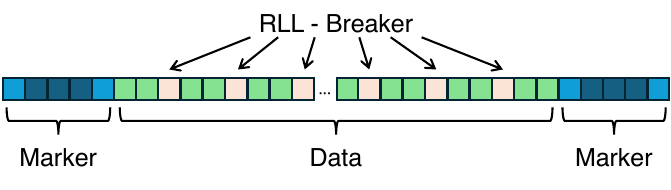}
  \caption{Example of a codeword with marker sequence and an RLL code}
  \label{fig:marker-code}
\end{figure}

\begin{construction}
  Let $\mathcal{X}$ be the set of all possible composite matrices over the base alphabet $q$ with resolution parameter $M$ and length $n$. We define the code $\mathcal{C} \subset \mathcal{X}$ as all composite matrices $\cmat{C}$ with entries $c_{i,j}$ where the conditions are fulfilled as in \autoref{ex:composite_marker} and \autoref{fig:marker-code}. 
  Formally written: 
  \begin{align*}
    &c_{i,1} = c_{i, \ell + 2} = c_{i,n - \ell - 1} = c_{i,n} = 
    \begin{cases}
      M & \text{ if } i = 2 \\
      0 & \text{ if } i \neq 2 
    \end{cases}
    \\
    &c_{i,j} = 
    \begin{cases}
      M & \text{ if } i = 1 \; \forall j \in [2,\ell+1] \cup [n - \ell, n -1] \\
      0 & \text{ if } i \neq 1 \; \forall j \in [2,\ell+1] \cup[n - \ell, n -1]
    \end{cases}
    \\
    &c_{1,j} = 0 \text{ if } j + 2 \mkern-10mu \mod \ell = 0 \; \forall  \ell + 3 \leq j \leq n - \ell - 2
  \end{align*}
\end{construction}

Given the composite strand break channel's behaviour, this code can reconstruct the data if sufficiently many strands have not experienced more than one break. An illustration of this can be found in \autoref{fig:marker-code}.

From the length of the marker and \eqref{eq:red-rll_triv}, we get the redundancy of the code $\mathcal{C}$ as 
\vspace{-0.2em}
\begin{align*}
  \red{\mathcal{C}} = 2 \ell + 4 + \left\lfloor \tfrac{n - 2(\ell +2)}{\ell} \right\rfloor \log_Q \big( \tfrac{Q}{Q-R} \big).
\end{align*}

Next, we want to determine the optimum value for $\ell$ by minimising the construction's redundancy. For simplicity, we will assume that $n$ is a multiple of $\ell$.

\begin{corollary}
  Given that $n$ is a multiple of $\ell$, the redundancy of the code $\mathcal{C}$ is minimized for 
  \vspace{-0.2em}
  \begin{align*}
    \ell^\ast  = \left( \tfrac{n-4}{2} \log_Q \big(\tfrac{Q}{Q-R} \big) \right)^\frac{1}{2}.
  \end{align*}
  This results in an overall redundancy of 
  \begin{align*}
    \red{\mathcal{C}} = 4 + 2 \left( 2 (n-4) \log_Q \big( \tfrac{Q}{Q-R} \big) \right)^{\frac{1}{2}} - 2 \log_Q \big( \tfrac{Q}{Q-R} \big).
  \end{align*}
\end{corollary}

\begin{IEEEproof}
  The results can be obtained by taking the derivative of the redundancy $\red{\mathcal{C}}$ with respect to $\ell$, setting it to zero and plugging it in.
\end{IEEEproof}

The following example illustrates the code construction for the 4-ary alphabet used in DNA.

\begin{example}
  Let $q = 4$, the resolution parameter $M = 6$, and therefore the number of composite symbols $Q = \binom{q+M-1}{q-1} = 84$. The number of unwanted symbols every $\ell$ positions in the data part is $R = Q - \binom{q-1 + M -1}{q-1-1} = 56$. Then we get the optimal marker length as $\ell \approx 0.24 \sqrt{n}$. The code is depicted in \autoref{fig:marker-code}.

  Assuming the marker consists of a run of $\mathrm{A}$'s, the \emph{RLL-Breaker} symbols appear every $\ell$ positions in the data part where the $\mathrm{A}$'s ratio is set to 0. Hence, there are only $Q - R = \binom{q-1 + M -1}{q-1-1} = 28$ symbols left for these positions.
\end{example}

Observe that the redundancy of an overall encoder, whose RLL-encoding component requires redundancy on the order of magnitude implied by \autoref{the:red_rll_low} and \autoref{the:red_rll_up}, is given by 
%\begin{align*}
  $\Theta\Big(\ell + \big(\tfrac{R}{Q}\big)^\ell\cdot n\Big)$.
%\end{align*}
As one addend is increasing and the other decreasing in~$\ell$, it is minimised for $\ell^\ast$ satisfying $\ell^\ast \big(\tfrac{Q}{R}\big)^{\ell^\ast} = \Theta(n)$, or equivalently 
\begin{align*}
  \ell^\ast = \log_{Q/R}\big(n/\log(n)\big) + O(1).
\end{align*}
This results in $\Theta(\log(n))$ redundant symbols. 

\section{Conclusion and Future Work}

This paper introduced a novel channel model to address strand breaks in composite DNA data storage. The channel model is designed to reflect realistic behaviour. We proposed a marker-based coding scheme to retrieve information when many strands experience a single break, aligning with experimental observations. Additionally, we generalised the concept of run-length-limited (RLL) codes for the composite setting and derived both lower and upper bounds on the redundancy of these codes. A practical code construction was presented, and the optimal marker length was determined. Future work could focus on improving the RLL encoder to narrow the gap to the theoretical bounds. Moreover, the code construction could be extended to handle multiple breaks or other error types such as insertions, deletions, or substitutions. Perhaps most importantly, the capacity of the composite channel with strand breaks itself should be ascertained.

\printbibliography

\newpage

%\section*{Appendix}
\appendix[Change of Summation in \eqref{eq:change_summation}]
We have the double summation
\begin{align*}
  \sum_{j=1}^{\ell-1} & \sum_{i=0}^{\ell-j-1} \left( \tfrac{R}{Q} \right)^i
  \\
  = & \quad \left( \tfrac{R}{Q} \right)^0 + \left( \tfrac{R}{Q} \right)^1 + \ldots + \left( \tfrac{R}{Q} \right)^{l-3} + \left( \tfrac{R}{Q} \right)^{l-2} \\
  &+ \left( \tfrac{R}{Q} \right)^0 + \left( \tfrac{R}{Q} \right)^1 + \ldots + \left( \tfrac{R}{Q} \right)^{l-3} \\
  & \mkern9mu \vdots \\
  &+ \left( \tfrac{R}{Q} \right)^0 + \left( \tfrac{R}{Q} \right)^1 \\
  &+  \left( \tfrac{R}{Q} \right)^0 \\
  = &\quad \left( \tfrac{R}{Q} \right)^0 + \left( \tfrac{R}{Q} \right)^0 + \ldots + \left( \tfrac{R}{Q} \right)^0 \text{\{ $l-2$ times\} } \\
  & + \left( \tfrac{R}{Q} \right)^1 + \left( \tfrac{R}{Q} \right)^1 + \ldots + \left( \tfrac{R}{Q} \right)^1 \text{\{ $l-3$ times\} } \\
  & \mkern9mu \vdots
  \\
  & + \left( \tfrac{R}{Q} \right)^{l-3} + \left( \tfrac{R}{Q} \right)^{l-3} \\
  & + \left( \tfrac{R}{Q} \right)^{l-2} \\
  = &\quad \sum_{i=0}^{l-2} \sum_{j=1}^{l-1-i} \left( \tfrac{R}{Q} \right)^i
\end{align*}

\end{document}